\documentstyle[12pt]{article}
\begin{document}
\def\BbbZ{{\rm Z}\!\!{\rm Z}}
\def\BbbR{{\rm I}\!{\rm R}}
\thispagestyle{empty}
\begin{center}

\null
\vskip-1truecm
\rightline{IC/95/399}
\vskip1truecm
International Atomic Energy Agency\\
and\\
United Nations Educational Scientific and Cultural Organization\\
\medskip
INTERNATIONAL CENTRE FOR THEORETICAL PHYSICS\\
\vskip2truecm
{\bf ON THE OVERLAP PRESCRIPTION\\ FOR LATTICE REGULARIZATION OF CHIRAL
FERMIONS\\}
\vskip2truecm
S. Randjbar--Daemi\quad and\quad J. Strathdee\\
International Centre for Theoretical Physics, Trieste, Italy.\\
\end{center}
\vskip1truecm
\centerline{ABSTRACT}
\baselineskip=24pt
\bigskip

Feynman rules for the vacuum amplitude of fermions coupled to external gauge
and Higgs fields in a domain wall lattice model are derived using
time--dependent perturbation theory. They have a clear and simple structure
corresponding to 1--loop vacuum graphs. Their continuum approximations are
extracted by isolating the infrared singularities and it is shown that, in each
order, they reduce to vacuum contributions for chiral fermions. In this sense
the lattice model is seen to constitute a valid regularization of the continuum
theory of chiral fermions coupled to weak and slowly varying gauge and Higgs
fields. The overlap amplitude, while not gauge invariant, exhibits a well
defined (modulo phase conventions) response to gauge transformations of the
background fields. This response reduces in the continuum limit to the expected
chiral anomaly, independently of the phase conventions.
\vskip1truecm
\begin{center}
{MIRAMARE -- TRIESTE\\
December 1995\\}
\end{center}

\newpage

\section{Introduction}

Recent efforts to provide a lattice regularized model for chiral fermions have
made some progress using Kaplan's idea of the domain wall [1]. This can be
formulated as a 4--dimensional Euclidean lattice embedded in 4+1--dimensional
spacetime. The continuous time coordinate is an artificial variable whose
purpose is to accomodate a barrier, or domain wall, with which 4--dimensional
chiral fermions can be associated. The barrier corresponds to a mass term
discontinuity in the time--dependent Hamiltonian of the 4+1--dimensional
system. Couplings to time--independent (external) gauge and Higgs fields are
included in this Hamiltonian. The aim is to compute the vacuum transition
amplitude for this 4+1--dimensional system and extract from it a quantity that
can be interpreted as the Euclidean vacuum amplitude for chiral fermions on a
4--dimensional lattice.

An equivalent picture, developed by Narayanan and Neuberger [2,3], interprets
the 4--dimensional Euclidean amplitude as the overlap of ground states
belonging to the two distinct Hamiltonians that govern the 4+1--dimensional
system on either side of the barrier.

The motivation behind these efforts is to obtain a model suitable for
numerical, i.e., non--perturbative, studies of chiral theory. It is not yet
clear that this aim will be achieved, but some encouraging results have been
obtained by Narayanan and Neuberger [3,4] in the context of a 2--dimensional
model whose continuum version is soluble.

Our purpose here is more limited. We want to show that the domain wall or
overlap prescription is perturbatively correct. By this we mean that, for weak
and slowly varying external gauge and Higgs fields, the prescription yields, in
each order, a regulated version of the Feynman graph that one expects to find
in a continuum theory of chiral fermions. Some work in this direction is
already available in the literature. On the one hand, a set of rules for
computing the perturbative contributions to the overlap amplitude was
developed by Neuberger and Narayanan [3]. On the other hand, the continuum
limit of low order contributions has been examined by several groups [2,5,6].
The latter work is concerned mainly with establishing that the expected chiral
anomalies are indeed present. Recently, however, we verified that the vacuum
polarization tensor for chiral fermions of the $SU(2)\times U(1)$ standard
model is recovered in the continuum limit [7]. Kaplan and Schmaltz [8] have
shown in the continuum version, that the phase of the overlap coincides with
the $\eta$--invariant of Alvarez--Gaum\'e {\em et al}. [9]. This is a
non--perturbative result. Some non--perturbative analytic work in 2--dimensions
is also available [10].

To set up the lattice model that corresponds to a collection of Weyl fermions
coupled to background gauge and Higgs fields in a 4--dimensional Euclidean
spacetime, one begins by doubling the number of fermion components.
Corresponding to each 2--component Weyl fermion, $\psi_L(x)$ or $\psi_R(x)$,
introduce a 4--component Dirac field, $\psi (n,t)$, defined on the sites of a
4--dimensional integer lattice, $n^\mu\in\BbbZ^4$. The ``time'' coordinate,
$t$, is continuous. Construct the Hamiltonian as a bilinear form,
\setcounter{equation}{0}
\renewcommand{\theequation}{1.\arabic{equation}}
\begin{equation}
H(t)=\sum_{n,m}\ \psi (n,t)^\dagger\ H(n,m,t)\ \psi (m,t)
\end{equation}
where the matrices $H(n,m,t)$ are covariant functionals of the background
fields. These background fields, $A_\mu (x)$ and $\phi (x)$, are assumed to be
smooth and independent of $t$. The functional dependence of the matrices
$H(n,m,t)$ is specified in detail in Appendix A. Here, we remark only that
their time--dependence is confined to the mass--like term,
\begin{equation}
\varepsilon (t)\Lambda\gamma_5T_c\delta_{nm}
\end{equation}
where $\varepsilon (t) =sign(t)$, $\Lambda$ is a positive parameter
representing the height of the domain wall and $T_c$ is a diagonal matrix with
eigenvalues $+1(-1)$ corresponding to right(left) handed Weyl fermions. It
commutes with the Dirac algebra. See Appendix A for details.

The structure of the Hamiltonian (1.1) is very simple. It has a discontinuity
at $t=0$ but is otherwise independent of time,
\begin{equation}
H(t)=\left\{\matrix{ H_+(A),&\quad t>0\cr \cr
H_-(A),&\quad t<0\cr}\right.
\end{equation}
where the argument, $A$, stands collectively for the background gauge and Higgs
fields. Although the Hamiltonian is discontinuous, the Heisenberg--picture
field, $\psi (n,t)$, is continuous at $t=0$ where it coincides with the
Schroedinger--picture field, $\psi (n)$. When the background fields are weak,
as we shall assume, there is a natural separation into free and interaction
terms,
\begin{equation}
H(t)=H_0(t)+V
\end{equation}
and one can set up the usual perturbation series. The only unusual feature here
is the discontinuity in the free Hamiltonian.

With the two Hamiltonians, $H_+(A)$ and $H_-(A)$, one can construct two
distinct normalized ground states, $\vert A+\rangle$ and $\vert A-\rangle$,
respectively. Of particular interest is the functional $\Gamma (A)$ defined by
the overlap,
\begin{equation}
\langle A+\vert A-\rangle =\langle +\vert -\rangle\ e^{-\Gamma (A)}
\end{equation}
where $\vert\pm\rangle$ denotes the respective ground states of the free
Hamiltonians, $H_{0\pm}=H_\pm (0)$. This is the functional whose perturbation
development we shall consider and which, we shall show, reduces in the
continuum limit to the connected vacuum amplitude for a set of Euclidean Weyl
fermions. An efficient way to compute $\Gamma (A)$ is by means of
time--dependent perturbation theory. The rules for expressing the contributions
in terms of connected 1--loop vacuum graphs are obtained in Sec.2.

Since the free Hamiltonian depends on time, the free particle Green's function
will not be invariant under time translations and, as a result, the detailed
structure of the perturbative contributions to $\Gamma (A)$ is more complicated
than in familiar theories. However, these complications tend to become
unimportant in the continuum limit. When the 4--momenta carried by external
fields are small compared to barrier height and inverse lattice spacing,
amplitudes are dominated by infrared singularities, i.e. thresholds associated
with the propagation of light fermions. The leading infrared singularities are
insensitive to lattice structure and can be computed by an effective continuum
field theory of chiral fermions in 4--dimensional Euclidean spacetime. It is
precisely this continuum theory that the lattice model regulates. The emergence
of a chiral continuum theory is discussed in Sec. 4. It depends crucially on
the infrared behaviour of the free fermion Green's function whose detailed
structure is considered in Appendix B.

The functional $\Gamma (A)$ is not gauge invariant. Under a gauge
transformation of the background fields, $A\to A^\theta$, it responds according
to
\begin{equation}
\Gamma (A^\theta )=\Gamma (A) -i\ \Phi_+(\theta ,A) +i\ \Phi_-(\theta ,A)
\end{equation}
where $\Phi_\pm$ are real angles associated with transformations of the ground
states, $\vert A\pm\rangle$. There is some arbitrariness in these angles that
reflects the role of phase conventions in the construction of $\vert
A\pm\rangle$. One possibility would be to impose the Brillouin--Wigner
convention: that the overlaps $\langle\pm\vert A\pm\rangle$ shall be real and
positive. This choice was adopted by Narayanan and Neuberger in their original
formulation of the overlap prescription [2]. It is well adapted to
time--independent perturbation theory and was used also in our low order
computations [5,7,11]. Here we shall use another convention that is better
adapted to time--dependent perturbation theory, to be explained in Sec.2. With
either of these conventions the difference, $\Phi_+-\Phi_-$, is non--vanishing
in general. In both of them, however, it can be shown that $\Phi_+-\Phi_-$
reduces to the standard chiral anomaly in the continuum limit. This means, in
particular, that $\Gamma (A)$ becomes gauge invariant in the continuum limit if
the Weyl fermions belong to an anomaly--free combination [5].

Gauge transformations are discussed in Sec.3 and the angles $\Phi_\pm$ are
defined there. The relative phase between the ground states of this paper and
the Brillouin--Wigner states used in earlier work is discussed in Appendix C
where the time--independent formalism is briefly reviewed.

A subtle point concerning gauge transformations and the continuum limit is
raised in Sec.4. This limit exhibits a lack of ``uniformity''. One finds that
the gauge variation of the continuum limit of the effective action differs from
the continuum limit of the gauge variation, $\Phi_+-\Phi_-$. This is because
the
continuum limit of $\Gamma (A)$ is dominated by infrared singularities that are
not present in $\Phi_+-\Phi_-$. The latter quantity, it will be seen, is
determined by massive fermions and reduces to a local form, the integral over
4--dimensional Euclidean spacetime of a pseudoscalar density in the slowly
varying gauge fields and their derivatives. The coefficients in this density
are finite lattice--dependent quantities. The chiral anomaly is a sub--dominant
effect from the infrared point of view. (This is only to be expected since, in
continuum gauge theory the anomaly arises in a parity violating amplitude which
is ultraviolet convergent and unambiguous, but whose gauge variation is, at
least superficially, ultraviolet divergent.) In order to recover the standard
consistent anomaly of continuum gauge theory from the local expression for
$\Phi_+-\Phi_-$ it is necessary to let the barrier height, $\Lambda$, become
vanishingly small relative to the lattice cutoff, $a^{-1}$. The condition,
$\Lambda a\ll 1$, was used in Ref.[5] where we obtained the chiral anomaly by
computing $\Phi_+-\Phi_-$. This condition is used implicitly in many studies of
the overlap prescription -- so called continuum models -- where the fermions
are represented by smooth fields in 5--dimensional spacetime, and ultraviolet
questions are ignored [2,3,11,12,8]. However, it should be recognized as a
non--essential technicality. In Sec.4 and Appendix B we show that the continuum
theory emerges as the infrared dominant part of $\Gamma (A)$ provided only that
the background fields are slowly varying on the scale of $\Lambda^{-1}$. It is
not necessary to assume $\Lambda^{-1}\gg a$.\footnote{We wish to
acknowledge a useful correspondence with H. Neuberger who insisted on this
point.}

In this paper we are exclusively concerned with the domain--wall--overlap
formulation of chiral gauge theories on the lattice. For some other approaches
see [13--17].

\section{Perturbation theory}

Since the domain wall problem is unusual in having a time dependent free
Hamiltonian we begin with a brief description of perturbation theory in the
interaction picture. We assume that the free Hamiltonian is invariant with
respect to lattice translations so that Fourier transforms can be used. The
interaction picture equations of motion take the form
\setcounter{equation}{0}
\renewcommand{\theequation}{2.\arabic{equation}}
\begin{eqnarray}
i\ \partial_t\ \psi (p,t ) &=& [\psi (p,t), H_0(t)]\nonumber\\
&=& H(p,t)\psi (p,t)
\end{eqnarray}
where $H(p,t)$ is an hermitian matrix, discontinuous at $t=0$ but otherwise
independent of time (see Appendix A). The solution of (2.1), continuous at
$t=0$, is given by
\begin{equation}
\psi (p,t)=\left\{\matrix{ e^{-itH_+(p)}\ \psi (p),&\quad t>0\cr
\cr
e^{-itH_-(p)}\ \psi (p),&\quad t<0\cr}\right.
\end{equation}
The time--dependent states of the interaction picture are governed by the usual
unitary operator,
\begin{equation}
\Omega (t) =\left\{\matrix{
T\left( e^{-i\int^t_0dt'V(t')}\right) ,&\quad t>0\cr \cr
\bar T\left( e^{-i\int^t_0dt'V(t')}\right) ,&\quad t<0\cr}\right.
\end{equation}
where $\bar T$ denotes antichronological ordering. In order that these
integrals converge for $t\to\pm\infty$, the operator $V(t)$ should include the
damping factor, $e^{-\varepsilon\vert t\vert}$, i.e.
\begin{equation}
i\ \partial_t\ V(t)=[V(t),H_0(t)]-i\varepsilon\ sgn(t)\ V(t)
\end{equation}

There are two free fermion ground states, $\vert +\rangle$ and $\vert
-\rangle$, defined by
\begin{equation}
H_{0+}\vert +\rangle =0,\qquad H_{0-}\vert -\rangle =0
\end{equation}
where the respective Dirac seas are filled. Expressions for the 1--body
Hamiltonians $H_\pm (p)$ are given in Appendix A. The details are not important
for now, except for the existence of a gap,
$$
\vert H_\pm (p)\vert\geq\omega_{\min}(\Lambda )>0
$$
where $\Lambda$ represents the height of the domain wall. Ground states of the
interacting theory can be generated adiabatically from these states [18].
Thus, for small but finite $\varepsilon$ define the asymptotic states
\begin{eqnarray}
\vert in\ \pm\rangle &=&\Omega (-\infty )^{-1}\vert\pm\rangle\nonumber\\
\vert out\ \pm\rangle &=&\Omega (\infty )^{-1}\vert\pm\rangle
\end{eqnarray}
When $\varepsilon$ tends to zero these states converge, apart from a singular
phase, onto eigenstates of the Schroedinger--picture Hamiltonians,
\begin{equation}
H_\pm (A) =H_{0\pm}+V
\end{equation}
where $A$ denotes a collection of time--independent external fields. One can
show [18]
\begin{eqnarray}
\lim_{\varepsilon\to 0}\left(\vert in\ \pm\rangle\ e^{-\hat E_\pm
/i\varepsilon}\right) &=&\vert A\pm\rangle\nonumber\\
\lim_{\varepsilon\to 0}\left(\vert out\ \pm\rangle\ e^{\hat E_\pm
/i\varepsilon}\right) &=&\vert A\pm\rangle
\end{eqnarray}
where the functionals $\hat E_\pm (A)$ are related to the ground state energies
defined by
\begin{equation}
H_\pm (A)\vert A\pm\rangle =\vert A\pm\rangle\ E_\pm (A)
\end{equation}
The relation between $\hat E$ and $E$ takes a simple form when both are
expanded
in powers of $V$, viz.
\begin{eqnarray}
E &=& E_1+E_2+E_3+\dots\nonumber\\
\hat E &=& E_1+{1\over 2}E_2+{1\over 3}E_3+\dots
\end{eqnarray}

To compute these energies using time--dependent perturbation theory one writes,
for example,
\begin{eqnarray*}
\langle out+\vert in+\rangle &=& e^{2\hat E_+/i\varepsilon}\\
&=&\langle +\vert\Omega (\infty )\ \Omega (-\infty )^{-1}\vert +\rangle\\
&=&\langle +\vert T\left( e^{-i\int^\infty_{-\infty}dt\ V(t)}\right)\vert
+\rangle
\end{eqnarray*}
which can be reduced to the computation of connected vacuum graphs,
\begin{equation}
\hat E_+=\lim_{\varepsilon\to 0}\ {i\varepsilon\over 2}\ \langle +\vert T
\left( e^{-i\int^\infty_{-\infty}dt\ V(t)}\right)\vert
+\rangle_{con}
\end{equation}
Since the interaction is bilinear,
\begin{equation}
V(t)=\int\left({dp\over 2\pi}\right)^4\left({dq\over 2\pi}\right)^4\ \psi
(p,t)^\dagger\ V(p,q)\ \psi (q,t)\ e^{-\varepsilon\vert t\vert}
\end{equation}
There is only one connected graph in each order. One finds,
\begin{eqnarray}
\hat E_+ &=&-{i\varepsilon\over 2}\ \sum_N\ {(-i)^N\over
N}\int^\infty_{-\infty} dt_1\dots
dt_N\ e^{-\varepsilon (\vert t_1\vert +\dots +\vert t_N\vert )}\cdot\nonumber\\
&&\cdot\int\left({dp_1\over 2\pi}\right)^4\dots\left({dp_N\over 2\pi}\right)^4\
tr\Big[ V(p_1,p_2)S_+(p_2,t_1-t_2)V(p_2,p_3)\dots\Bigr.\nonumber\\
&&\hspace{5cm}\Bigl. \dots V(p_N,p_1)S_+(p_1,t_N-t_1)\Bigr]
\end{eqnarray}
where the propagator $S_+$ is defined by
\begin{equation}
\langle +\vert T(\psi (q,t)\ \psi (p,t')^\dagger )\vert +\rangle =(2\pi
)^4\delta_{2\pi}(q-p)\ S_+(q,t-t')
\end{equation}
It is invariant with respect to time translations because the time dependence
of $\psi (q,t)$ is determined here by the time--independent Hamiltonian,
$H_{0+}$. Indeed, we can write
\begin{equation}
S_+(q,t-t')=\int^\infty_{-\infty} {dE\over 2\pi}\ {i\over E-H_+(q)+i\eta\
sgn(H_+)}\ e^{-iE(t-t')}
\end{equation}
with $\eta >0$. The momentum integrations in (2.13) are compact. They range
over the Brillouin zone, a cell of volume $(2\pi )^4$ in lattice units. The
delta function in (2.14) is periodic with respect to reciprocal lattice
translations.

An expression analogous to (2.13) can be written for $\hat E_-$ by replacing
$S_+$ with $S_-$. More interesting is the transition amplitude that includes
the barrier effect,
\begin{equation}
\langle out +\vert in -\rangle =\langle +\vert T\left(
e^{-i\int^\infty_{-\infty}dt\ V(t)}\right) \vert -\rangle
\end{equation}
where the time dependence of $\psi (q,t)$ is now determined by the
time--dependent Hamiltonian, $H_0(t)$. Here one computes the connected vacuum
graphs using the propagator $S_F$, defined by
\begin{equation}
\langle +\vert T(\psi (q,t)\ \psi (p,t')^\dagger )\vert -\rangle =
\langle +\vert -\rangle\ (2\pi
)^4\delta_{2\pi}(q-p)\ S_F(q,t,t')
\end{equation}
This propagator is not invariant with respect to time translations. It is very
much more complicated than (2.15) and it includes long range effects due to the
propagation of chiral fermions. Its detailed structure is discussed in Appendix
B.

The order $N$ contribution to $-\langle out +\vert in-\rangle_{con}$ is given
by an expression analogous to (2.13),
\begin{eqnarray}
&&{(-i)^N\over N}\int dt_1\dots dt_N\ e^{-\varepsilon (\vert t_1\vert +\dots
+\vert t_N\vert )}\cdot\nonumber\\
&&\cdot\int\left({dp_1\over 2\pi}\right)^4\dots\left({dp_N\over 2\pi}\right)^4\
tr\Bigl[V(p_1,p_2)S_F(p_2,t_1,t_2)V(p_2,p_3)\dots\Bigr.\nonumber\\
&&\hspace{5cm}\Bigl.\dots V(p_N,p_1)S_F(p_1,t_N,t_1)\Bigr]
\end{eqnarray}

Finally, these results can be put together to give the effective action
functional, $\Gamma (A)$, defined by (1.5),
\begin{eqnarray*}
\langle +\vert -\rangle\ e^{-\Gamma (A)} &=& \langle A+\vert A-\rangle\\
&=& \langle out +\vert in-\rangle\ e^{-(\hat E_++\hat E_-)/i\varepsilon}\\
&=&\langle out +\vert in-\rangle \langle out +\vert in+\rangle^{-1/2}\ \langle
out -\vert in-\rangle^{-1/2}
\end{eqnarray*}
or, in terms of connected vacuum graphs,
\begin{equation}
\Gamma (A) =-\langle out +\vert in-\rangle_{con} +{1\over 2}\
\langle out +\vert in+\rangle_{con} +{1\over 2}\
\langle out -\vert in-\rangle_{con}
\end{equation}
The respective terms are to be computed using the propagators, $S_F,S_+$ and
$S_-$. The limit $\varepsilon\to 0$ is understood. In fact, the auxiliary
pieces, $\langle out \pm\vert in\pm\rangle_{con}$, are needed only to cancel
the singularity in $\langle out +\vert in-\rangle_{con}$ as $\varepsilon$ tends
to zero. The effective action is obtained as the regular part of $-\langle out
+\vert in-\rangle_{con}$.

\section{Gauge transformations}

Infinitesimal time--independent gauge transformations are generated by the
operator
\setcounter{equation}{0}
\renewcommand{\theequation}{3.\arabic{equation}}
\begin{eqnarray}
F_\theta &=&\Sigma\ \psi (n)^\dagger\ \theta (n)\ \psi (n)\nonumber\\
&=& \int\left({dp\over 2\pi}\right)^4\left({dq\over 2\pi}\right)^4\
\psi (p)^\dagger\ \tilde\theta (p-q)\ \psi (q)
\end{eqnarray}
where $\theta (n)$ is an  hermitian matrix belonging to the algebra of the
gauge
group. It is slowly varying on the lattice and will be interpolated by a smooth
function, $\theta (x)$, that defines the transformations of the background
fields, $A\to A^\theta$, such that
\begin{equation}
e^{iF_\theta}\ H_\pm (A)\ e^{-iF_\theta} = H_\pm (A^\theta )
\end{equation}

Since the ground states are non--degenerate, at least in perturbation theory,
it follows that they must transform according to
\begin{equation}
e^{iF_\theta}\vert A\pm\rangle =\vert A^\theta\pm\rangle\ e^{i\Phi_\pm (\theta
,A)}
\end{equation}
where the angles $\Phi_\pm$ are real. These angles provide a representation of
the group. Thus, if the product of two group elements is defined by
$$
e^{i\theta_1}\ e^{i\theta_2} =e^{i\theta_{12}}
$$
then it is easy to show that the corresponding composition rule for $\Phi_\pm$
is given by
\begin{equation}
\Phi (\theta_1,A^{\theta_2})+\Phi (\theta_2,A)=\Phi (\theta_{12},A)
\end{equation}

Gauge transformations of the effective action, $\Gamma (A)$, are obtained by
substituting (3.3) into the definition (1.5),
\begin{equation}
\Gamma (A^\theta )=\Gamma (A) -i\ \Phi_+(\theta ,A)+i\ \Phi_-(\theta ,A)
\end{equation}
There is no reason to expect the difference, $\Phi_+-\Phi_-$, to vanish in
general but we should expect some simplifications to occur when the background
fields are slowly varying. To see what happens it is necessary to compute these
angles in perturbation theory.

To first order in $\theta$ the transformation rule (3.3) takes the form
$$
\delta_\theta\vert A\pm\rangle =i(F_\theta -\Phi_\pm )\vert A\pm\rangle
$$
which implies, for example,
\begin{equation}
\Phi_+={\langle +\vert F_\theta\vert A+\rangle\over\langle +\vert A+\rangle}
+i\ \delta_\theta\ \ell n\langle +\vert A+\rangle
\end{equation}
and likewise for $\Phi_-$. This can be calculated in time--dependent
perturbation theory using the adiabatic formula (2.8). Firstly,
\begin{eqnarray}
i\ \delta_\theta\ \ell n\langle +\vert A+\rangle &=& i\ \delta_\theta\ \ell
n\langle +\vert in+\rangle -{1\over\varepsilon}\ \delta_\theta\hat
E_+\nonumber\\
&=&\int^0_{-\infty} dt\ {\langle +\vert T\left(\delta_\theta V(t)\
e^{-i\int^0_{-\infty} dt'V(t')}\right)\vert +\rangle\over\langle +\vert in
+\rangle} -{1\over{\varepsilon}}\ \delta_\theta\hat E_+
\end{eqnarray}
A suitable formula for $\delta_\theta V(t)$ can be extracted from (3.2),
\begin{eqnarray*}
\delta_\theta V(t) &=&e^{itH_{0+}}\ \delta_\theta V\ e^{-itH_{0+}}\
e^{\varepsilon t}\\
&=& e^{itH_{0+}}\ i[F_\theta ,H_{0+}+V]\ e^{-itH_{0+}}\ e^{\varepsilon t}\\
&=& e^{\varepsilon t}\ i[F_\theta (t),H_{0+}]+i[F_\theta (t),V(t)]\\
&=& -e^{\varepsilon t}\ \partial_t\ F_\theta(t) +i[F_\theta (t),V(t)]
\end{eqnarray*}
which implies
\begin{eqnarray}
&&\langle +\vert T\Biggl(\delta_\theta
V(t)\ e^{-i\int^0_{-\infty}dt'V(t')}\Biggr)\vert +\rangle =\nonumber\\
&&\qquad=-e^{\varepsilon t}\ \partial_t\langle +\vert T\left( F_\theta (t)\
e^{-i\int^0_{-\infty}dt'V(t')}\right)\vert +\rangle\nonumber\\
&&\qquad\quad -(e^{\varepsilon t}-1)\langle +\vert T\left( i[F_\theta
(t),V(t)]e^{-i\int^0_{-\infty}dt' V(t')}\right)\vert +\rangle
\end{eqnarray}

For Abelian symmetries the result is relatively simple. In such cases the
energy $E_+(A)$ is invariant in each order so that $\delta_\theta\hat E_+=0$.
Also, the commutator $[F_\theta , V]$ vanishes and the second part of (3.8) is
absent. It follows that
\begin{eqnarray}
\Phi_+ &=&{\langle +\vert F_\theta\vert in+\rangle\over\langle +\vert
in+\rangle} -\lim_{\varepsilon\to 0}\int^0_{-\infty}dt\ e^{\varepsilon t}\
\partial_t\ {\langle +\vert T\left( F_\theta (t)e^{-i\int^0_{-\infty}dt'
V(t')}\right)\vert +\rangle\over\langle +\vert in+\rangle}\nonumber\\
&=&\lim_{\varepsilon\to 0}\ \varepsilon \int^0_{-\infty} dt\ e^{\varepsilon t}
\langle +\vert T\left( F_\theta (t)\ e^{-\int^0_{-\infty}dt'V(t')}\right)\vert
+\rangle_{con}
\end{eqnarray}

For non--Abelian symmetries the second part of (3.8) cannot be ignored. It
contributes a singular term that cancels $\delta_\theta\hat E/\varepsilon$ from
(3.7) and a regular term that must be retained. Hence we can write
\begin{equation}
\Phi_+=reg\int^0_{-\infty} dt \langle +\vert T\left(\left\{ \varepsilon\
e^{\varepsilon t} F_\theta (t)-(e^{\varepsilon t}-1)\ i[F_\theta
(t),V(t)]\right\} e^{-i\int^0_{-\infty} dt'V(t')}\right)\vert +\rangle
\end{equation}
meaning the regular part at $\varepsilon =0$. A comparison of this result with
the corresponding angle obtained in the time--independent formalism using the
Brillouin--Wigner phase convention is made in Appendix C where (3.10) is
computed up to second order in $V(0)$.

\section{Infrared behaviour}

Having obtained general expressions for the perturbative contributions to the
vacuum amplitude, $\langle out +\vert in-\rangle$, we now consider how to
approximate them when the external fields are slowly varying.

In each order, the formula (2.18) corresponds to a 1--loop vacuum graph
constructed from the vertices, $V(p,q)$, and propagators, $S_F(p,t,t')$. The
vertex is expressible as an expansion in powers of the gauge field, $A_\mu
(x)$, the first two terms of which are given by (A.12). In general it should
include powers of the Higgs field, $\phi (x)$, as well, but we are simplifying
the structure by choosing $\phi$ to be constant and incorporating it in the
free propagator, $S_F$. This propagator, discussed in Appendix B, is given by
(B.14). While its general structure is quite complicated, near $p=\phi =0$ it
simplifies to the form given by (B.15), which exhibits a pole corresponding to
the propagation of light chiral fermions.

The loop integration in (2.18) ranges over the Brillouin zone, a cell of volume
$(2\pi )^4$ in lattice units. There is, of course, no ultraviolet divergence.
What interests us here is the possibility of infrared divergences which we
expect to dominate the amplitude when the gauge fields are slowly varying and
the Higgs field is small. To see that there is an infrared singularity it is
sufficient to examine the integrand of (2.18) in the vicinity of the point,
$p_1=p_2=\dots =p_N=\phi =0$. In this region the vertex (A.12) reduces to
\setcounter{equation}{0}
\renewcommand{\theequation}{4.\arabic{equation}}
\begin{equation}
V(p,q)\simeq -i\ \gamma_5\gamma^\mu\tilde A_\mu (p-q)+\dots
\end{equation}
and the propagator is dominated by the pole term in (B.15),
\begin{equation}
S_F(p,t,t')\simeq {1+\gamma_5T_c\over 2}\ {1\over ip\!\!\!/ +\phi\cdot T}\
\gamma_5\Lambda\ e^{-i\Lambda (\vert t\vert +\vert t'\vert )}+\dots
\end{equation}
Integration over the time coordinates, $t_1,\dots ,t_N$, is trivial in this
approximation since
$$
\Lambda\int^\infty_{-\infty} dt\ e^{-(\varepsilon +2i\Lambda )\vert
t\vert}={2\Lambda\over\varepsilon +2i\Lambda}\to{1\over i}
$$
in the limit, $\varepsilon\to 0$. Hence the integrand of (2.18) reduces to
\begin{eqnarray}
{i^N\over N}\ tr\left[\tilde A\!\!\!/ (p_1-p_2)\ {1+\gamma_5T_c\over 2}\
{1\over
ip\!\!\!/_2+\phi\cdot T}\dots\tilde A\!\!\!/ (p_N-p_1){1+\gamma_5T_c\over
2} {1\over
ip\!\!\!/_1+\phi\cdot T}\right]\nonumber\\
={i^N\over N}\ tr\left[\tilde A\!\!\!/ (k_1)\ {1+\gamma_5T_c\over 2}\ {1\over
i(p\!\!\!/-k\!\!\!/_1)+\phi\cdot T}\dots\tilde A\!\!\!/
(k_N){1+\gamma_5T_c\over
2} {1\over ip\!\!\!/ +\phi\cdot T}\right]
\end{eqnarray}
where $p_1=p-k_1,p_2=p-k_1-k_2,\dots$ and the external momenta are constrained
to satisfy $k_1+k_2+\dots +k_N=0$. For $N\geq 4$ there is clearly an infrared
singularity at $k_1=\dots =k_N=\phi =0$, because the loop integration then
diverges at $p=0$. For $N\leq 3$, derivatives of order $3-N$ with respect to
external momenta also diverge. These singularities are of course threshold
effects associated with the propagation of light chiral fermions near their
mass shell. The expression (4.3) is what one would expect to find in a
4--dimensional continuum theory described by the (Euclidean) Lagrangian density
\begin{equation}
{\cal L}=\bar\psi (\partial\!\!\!/ -iA\!\!\!/ +\phi\cdot T)\
{1+\gamma_5T_c\over
2}\ \psi
\end{equation}
Our point is that (4.3) emerges as the dominant infrared effect in the lattice
model. Subleading terms, down with respect to (4.3) by powers of $k_1,\dots
,k_N,\phi$ could be computed by improving the approximate formulae (4.1) and
(4.2), but they would be lattice dependent and should be interpreted as scaling
violations, irrelevant in the continuum approximation. In this sense the
lattice
overlap, or domain wall, prescription constitutes an ultraviolet regularization
of the continuum system (4.4).

To summarize our approach: we seek to isolate the contributions that are
singular in the infrared. The leading singularity is lattice--independent,
sub--leading and non--singular quantities are sensitive to the lattice and
should not be computed. Lattice dependent quantities are either not relevant to
the continuum theory or they can be incorporated in counterterms. A detailed
exposition of this approach as applied to the vacuum polarization tensor
$(N=2)$ is given in Ref.[7].

An important qualification should be made. The expression (4.3) represents the
dominant infrared contribution only if the propagator $S_F(p,t,t')$ has no
other poles. The argument assumes that $p=0$ is the only point in the Brillouin
zone where the propagator is singular. It must be shown explicitly that there
are no other such points, i.e. no doubling of fermions [19,20]. This matter is
dealt with in Appendix B.

It may be remarked that, since the infrared dominant and lattice independent
contributions (4.3) coincide exactly with the continuum theory formulae, they
must also carry the expected chiral anomalies. For example, with $N=3$ and, for
simplicity, $\phi =0$, the parity violating part of the amplitude is given by
\begin{eqnarray}
&&\Gamma_5(A) =-{i^3\over 3}\int\left({dk_1\over 2\pi}\right)^4
\left({dk_2\over 2\pi}\right)^4\left({dk_3\over 2\pi}\right)^4(2\pi
)^4\ \delta_4(\Sigma k)\cdot\nonumber\\
&&\cdot\int\left({dp\over 2\pi}\right)^4\ tr\Biggl[{\gamma_5T_c\over 2}\ \tilde
A\!\!\!/(k_1)\ {1\over i(p\!\!\!/-k\!\!\!/_1)}\ \tilde A\!\!\!/(k_2)\ {1\over
i(p\!\!\!/-k\!\!\!/_1-k\!\!\!/_2)}\tilde A\!\!\!/(k_3)\ {1\over ip\!\!\!/}
\Biggr]
\end{eqnarray}
where the loop integration is understood to comprise a small region around
$p=0$. The asymptotically dominant contribution to $\Gamma_5$ when
$k_1,k_2,k_3$ tend to zero is obtainable, however, by extending the range of
$p$ to the entire $\BbbR^4$ since the resulting integral is, in fact,
ultraviolet convergent. (There is no $SO(4)$ invariant, pseudoscalar local term
of dimension 4 that could serve as a counterterm.) One may calculate this
amplitude and verify that it satisfies
\begin{equation}
\delta_\theta\ \Gamma_5(A)=-{i\over 24\pi^2}\int d^4x\
\varepsilon^{\kappa\lambda\mu\nu}\ tr\Big( T_c\ \partial_\kappa\
A_\lambda\ \partial_\mu\ A_\nu\ \theta \Bigr)
\end{equation}
to second order in $A$.

It is interesting to compare the result (4.6) with the general formula (3.5)
or, to first order in $\theta$,
\begin{equation}
\delta_\theta\ \Gamma (A) =-i\ \Phi_+(\theta ,A)+i\ \Phi_-(\theta ,A)
\end{equation}
The angle $\Phi_+$ is given by (3.10) which can be expanded in powers of the
interaction, $V$. The result up to terms of second order is given by (C.14). In
the time--dependent formalism the angles $\Phi_+$ and $\Phi_-$ are computed
using the Green's functions $S_+$ and $S_-$, respectively. These propagators,
discussed in Appendix B, do not have a long range structure. They are regular
at $p=0$,
\begin{equation}
S_\pm (p,t-t')={1\over 2}\Bigl(\varepsilon (t-t')\pm\gamma_5T_c\Bigr)\
e^{-i\Lambda\vert t-t'\vert}+O\left({p\over\Lambda}\right)
\end{equation}
This means that the functionals $\Phi_+$ and $\Phi_-$ do not have any
singularities. For slowly varying fields they are effectively local, i.e.
expressible as integrals over 4--dimensional Euclidean spacetime of local
functions of $A(x),\partial A(x),\dots$\ \ The coefficients in these local
functions are, of course, lattice dependent. Indeed, they must scale with the
barrier height, $\Lambda$, and lattice cutoff, $a^{-1}$, to a power given by
their canonical dimension. In the continuum limit, coefficients with negative
dimensionality will tend to zero and we may therefore restrict attention to
those with non--negative dimensionality. For the gauge variation (4.7) only the
pseudoscalar, $\Phi_+-\Phi_-$, needs to be considered and this functional
involves only one relevant quantity, the dimension zero coefficient of the
integral in (4.6). However, this coefficient generally depends in a complicated
way on $\Lambda a$ and other dimensionless lattice parameters. It is
expressible as an integral over the Brillouin zone and it does not agree with
the coefficient in (4.6). But one can show that agreement is recovered in the
limit, $\Lambda a\to 0$. (This calculation was carried out in Ref.[5] where it
was shown that the integral over the Brillouin zone develops an infrared
singularity at $p=0$ in the limit $\Lambda\to 0$.) This phenomenon seems to
indicate a lack of ``uniformity'' in the continuum limit. The gauge variation
of the continuum limit of $\Gamma (A)$ does not coincide with the continuum
limit of its gauge variation unless the secondary limit, $\Lambda\to 0$, is
also taken. It should be interpreted as the lattice version of an effect that
is familiar in continuum chiral theory. There, it is well known that the parity
violating amplitude is ultraviolet convergent and unambiguous, as must be its
gauge variation, the  (consistent) chiral  anomaly. On the other hand, this
gauge variation is
expressible as the difference of two formally identical, but ultraviolet
divergent integrals, that have to be regulated carefully in order to obtain the
correct anomaly. Ultraviolet convergent (divergent) integrals in continuum
theory correspond to infrared singular (non--singular) integrals in lattice
theory.

In obtaining the continuum limit of the overlap amplitude we have used the
approximate formula, (4.2), the leading term in an expansion of $S_F$ in powers
of $p/\Lambda$ and $ap$. The result (4.3) is presumably valid if $k_1,k_2,\dots
,\phi$ are all small compared to $\Lambda$ and $a^{-1}$. No condition on the
magnitude of $\Lambda a$ is involved. However, in view of the non--uniform
response to gauge transformations outlined above, we suspect that it would be
safer to choose $\Lambda a$ to be small.

\section{Conclusions}

In this paper we have provided a set of rules for computing an overlap
amplitude,
 order by order, in weak field aproximation.  We have shown that this amplitude
can be interpreted as a lattice regularization of the vacuum amplitude for a 4-
dimensional Euclidean continuum theory of chiral fermions coupled to
background gauge and Higgs
fields.

We find that the most efficient approach is through the use of
time-dependent perturbation
theory.  The 4-dimensional Euclidean lattice is embedded in 4+1-dimensional
Minkowski space with a
continuous and unbounded time coordinate.  This leads to an expansion of
the overlap in terms of
1-loop vacuum graphs.  We found it convenient in this work to use real time
formalism but
one could easily construct analogous formulae using imaginary time.  When
the background fields are
slowly varying on the lattice scale and also on the scale of the inverse
barrier height, the
perturbative expressions simplify.  The infrared dominant term can be
extracted in each order and,
after integrating the time coordinates, put into correspondence with the
continuum formula for that
order.

Although the lattice amplitude may not be itself gauge invariant, it is
guaranteed that the
continuum limit, i.e. the infrared dominant part, will be gauge invariant
up to chiral anomalies.
These would have to be compensated in the standard way in order to recover
a fully gauge
invariant theory in the continuum.

\section*{Acknowledgements}

 We gratefully acknowledge many useful discussions with R. Iengo and
correspondence with D. Kaplan,
R. Narayanan and H. Neuberger.

\newpage

\section*{Appendix A: The model}

The purpose of this Appendix is to specify the details of the lattice model
including the functions $H_\pm (p), V(p,q)$ used in Sec.2 [5].

The dynamical variables comprise a set of Dirac fields, $\psi _i(n,t),\ \
i=1,\dots ,N$ associated with the sites of a 4--dimensional integer lattice,
$n\in\BbbZ^4$. The time coordinate is continuous. The Hamiltonian is bilinear
and time--dependent,
\setcounter{equation}{0}
\renewcommand{\theequation}{A.\arabic{equation}}
\begin{equation}
H(t)=\sum_{n,m}\ \psi (n,t)^\dagger\ H(n,m,t)\ \psi (m,t)
\end{equation}
where the coefficients $H(n,m,t)$ are $4N\times 4N$ matrices acting in the
product of Dirac and flavour spaces. They incorporate the couplings to external
gauge and Higgs fields, $A_\mu (x)$ and $\phi (x)$, that are assumed to be
smooth functions on $\BbbR^4$, interpolating the lattice sites. The general
structure is
\begin{equation}
H(n,m,t)=H_0(n-m)\ U(n,m\vert A)+\delta_{nm}M(n,t\vert\phi )
\end{equation}
where $H_0,U$ and $M$ are matrices defined as follows.

Firstly, the gauge factor $U$ is specified by a path ordered exponential where
the path is chosen to be the straight line joining lattice sites $n$ and $m$,
\begin{equation}
U(n,m\vert A)=P\left(\exp\left\{ i\int^n_m dx^\mu\ A^\alpha_\mu
(x)T_\alpha\right\}\right)
\end{equation}
where $x^\mu (t)=tn^\mu +(1-t)m^\mu ,\ 0\leq t\leq 1$. The $N\times N$
hermitian matrices $T^\alpha$ belong to the algebra of the gauge group.

Next, the mass term, $M$, includes the Higgs background and the barrier effect,
\begin{equation}
M(n,t\vert\phi )=\left\{\matrix{
\gamma_5(\phi^i(n)T_i+\Lambda\ T_c), &\quad t>0\cr \cr
\gamma_5(\phi^i(n)T_i-\Lambda\ T_c), &\quad t<0\cr}\right.
\end{equation}
The $N\times N$ matrices $T_i$ incorporate Yukawa coupling parameters and
define the representation of the gauge group to which $\phi^i$ belongs, i.e.
\begin{equation}
[T_i,T_\alpha ]=i(t_\alpha )_i\ ^j\ T_j
\end{equation}
The ``chirality'' matrix, $T_c$, is diagonal with eigenvalues $\pm 1$
corresponding to right or left handed flavours in the continuum limit. This
matrix is required to be gauge invariant,
\begin{equation}
[T_c,T_\alpha ]=0
\end{equation}
On the other hand, since the role of Higgs fields is to connect left with right
handed fermions, the matrices $T_i$ are required to anticommute with $T_c$,
\begin{equation}
\{ T_c,T_i\} =0
\end{equation}

Finally, to specify the hopping term $H_0(n-m)$, it is useful to employ Fourier
series. Define the Fourier components, $\psi (p,t)$, by the lattice sum
$$
\psi (p,t)=\sum_n\ \psi (n,t)\ e^{-ipn}
$$
where $pn=p_\mu n^\mu$. These components are periodic in momenta with period
$2\pi$ (in lattice units). The translation invariant hopping term is
represented by the Fourier integral
\begin{equation}
H_0(n-m)=\int_{BZ} \left({dp\over 2\pi}\right)^4\ \gamma_5(i\gamma^\mu\ C_\mu
(p)+B(p)\ T_c)\ e^{ip(n-m)}
\end{equation}
where the integral ranges over a Brillouin zone, a cell of volume $(2\pi )^4$
in lattice units. The functions $C_\mu$ and $B$ are real and periodic. Their
detailed structure is not important for us, except in the infrared. We require
that they have no common zeroes, apart from the origin. Near $p=0$ they must
take the form,
\begin{equation}
C_\mu (p)\simeq p_\mu +\dots ,\quad B(p)\simeq r\ p^2+\dots
\end{equation}
where $p^2=g^{\mu\nu}p_\mu p_\nu$, and $r$ is a constant (Wilson parameter)
[19]. The metric tensor, $g^{\mu\nu}$, is Euclidean and we may suppose that it
is invariant with respect to one of the 4--dimensional crystal groups. This
tensor is involved also in the Dirac algebra,
\begin{equation}
\{\gamma^\mu ,\gamma^\nu\} =2g^{\mu\nu}
\end{equation}
It is convenient to normalize the metric such that $\det g=1$. In choosing a
metric tensor we are essentially choosing a crystal structure for the lattice.
The particular choice is presumably not very important as regards infrared
behaviour, although we should insist that the invariance group of $C_\mu$ and
$B$ be large enough to enforce the structure (A.9) near $p=0$. We also assume
that the crystal symmetry includes reflections with the barrier term
transforming as a pseudoscalar. This will ensure that quantities such as
$\Phi_+-\Phi_-$ transform as pseudoscalars under space reflections of the
background fields when $\Lambda^{-1}\gg a$ and lattice effects are ignored
i.e., in the infrared regime.

Under gauge transformations,
\begin{eqnarray*}
A(x)\to A^\theta (x) &=& e^{i\theta (x)}\ (A(x)+id)\ e^{-i\theta (x)}\\
\phi (x)\cdot T\to\phi^\theta (x)\cdot T &=& e^{i\theta (x)}\ \phi (x)\cdot T\
e^{-i\theta (x)}
\end{eqnarray*}
the coefficient matrices (A.2) are clearly covariant,
$$
H(n,m,t)\to e^{i\theta (n)}\ H(n,m,t)\ e^{-i\theta (m)}
$$
since $\theta (x)=\theta^\alpha (x) T_\alpha$ commutes with $T_c$. This
guarantees the formula (3.2). (In the text we have used the collective
notation, $A$, to represent both gauge and Higgs fields.)

In developing the perturbative formulae of Sec.2 we have assumed that the Higgs
field is constant and incorporated it into the free Hamiltonian. The free
Hamiltonians, $H_{0\pm}$, are therefore defined by the 1--body expressions,
\begin{equation}
H_\pm (p)=\gamma_5\Big( i\ \gamma^\mu\ C(p)+\phi\cdot T+(B(p)\pm\Lambda
)T_c\Bigr)
\end{equation}
The vertex functions are obtained by expanding (A.3) in powers of the gauge
field. One obtains,
\begin{eqnarray}
V(p,p) &=&-\int\left({dk\over 2\pi}\right)^4\ \tilde A^\alpha_\mu (k) \
(2\pi)^4\ \delta_{2\pi}(-p+q+k)\int^1_0dt\ H(p-tk)^{,\mu}\
T_\alpha +\nonumber\\
&&+{1\over 2}\int\left({dk_1\over 2\pi}\right)^4\left({dk_2\over
2\pi}\right)^4\ \tilde A^\alpha_\mu (k_1)\ \tilde A^\beta_\nu (k_2)\ (2\pi)^4\
\delta_{2\pi} (-p+q+k_1+k_2)\cdot\nonumber\\
&&\cdot \int^1_0 dt_1dt_2\ H(p-t_1k_1-t_2k_2)^{,\mu\nu}\Bigl(\theta
(t_1-t_2)T_\alpha T_\beta +\theta (t_2-t_1)T_\beta T_\alpha\Bigr)\nonumber \\
&&+\dots
\end{eqnarray}
where $H(p)^{,\mu}=\partial H_\pm (p)/\partial p_\mu$, etc. The periodic delta
function is defined by the lattice sum,
\begin{eqnarray*}
(2\pi )^4\ \delta_{2\pi}(p) &=& \sum_n\ e^{ipn}\\
&=&\sum_n\ (2\pi )^4\ \delta_4(p+2\pi n)
\end{eqnarray*}
The momentum integrals in (A.12) are over $\BbbR^4$ but, since $A_\mu (x)$ is
assumed to be slowly varying, its Fourier transform $\tilde A_\mu (k)$ is
concentrated around $k=0$.

The only part of (A.12) that is needed in the continuum limit is the first term
near $p=q=0$,
\begin{equation}
V(p,q)=-i\ \gamma_5\gamma^\mu\ \tilde A^\alpha_\mu (p-q)\ T_\alpha +\dots
\end{equation}

\newpage

\section*{Appendix B: Free fermions}

The purpose of this Appendix is to examine the spectrum of free fermion states
and derive expressions for the propagators, $S_F,S_+$ and $S_-$.

The square of the 1--body Hamiltonian (A.11) is proportional to the unit Dirac
matrix,
\setcounter{equation}{0}
\renewcommand{\theequation}{B.\arabic{equation}}
\begin{equation}
H_\pm (p)^2=g^{\mu\nu}\ C_\mu C_\nu +(\phi\cdot T)^2+(B\pm \Lambda )^2
\end{equation}
since $T^2_c=1$ and $\{ T_i,T_c\} =0$. Choose a set of $2N$ orthonormal spinors
$\chi (\sigma )$ such that [7]
\begin{eqnarray}
\gamma_5T_c\ \chi (\sigma ) &=&\chi (\sigma ),\quad\qquad \sigma =1,\dots
,2N\nonumber\\
(\phi\cdot T)^2\ \chi (\sigma ) &=& m^2_\sigma \ \chi (\sigma )
\end{eqnarray}
and define the eigenspinors of $H_\pm (p)$,
\begin{eqnarray}
u_\pm (p,\sigma ) &=& {\omega_\pm +H_\pm (p)\over
\sqrt{2\omega_\pm (\omega_\pm +B\pm\Lambda )}}\ \chi (\sigma )\nonumber \\
v_\pm (p,\sigma ) &=& {\omega_\pm -H_\pm (p)\over
\sqrt{2\omega_\pm (\omega_\pm -B\mp\Lambda )}}\ \chi (\sigma )
\end{eqnarray}
where $\omega_\pm (p,\sigma )$ is given by the positive square root,
\begin{equation}
\omega_\pm (p,\sigma ) =\sqrt{ C(p)^2+m^2_\sigma +(B(p)\pm\Lambda )^2}
\end{equation}
There are no zero modes, even in the absence of Higgs fields $(m_\sigma =0)$.
The positive and negative energy eigenspinors, $u_+,v_+$ of $H_+(p)$ comprise a
complete, orthonormal set. Likewise for $u_-$ and $v_-$. The two sets are
related by a unitary transformation,
\begin{eqnarray}
u_- &=& u_+\cos\beta- v_+\sin\beta\nonumber \\
v_- &=& u_+\sin\beta+ v_+\cos\beta
\end{eqnarray}
where the angle $\beta (p,\sigma )$ can be chosen to lie in the interval
$(0,\pi /2)$. It is given by
\begin{equation}
\cos\beta = \sqrt{{\omega_+-B-\Lambda\over 2\omega_+}\
{\omega_--B+\Lambda\over 2\omega_-}} +
\sqrt{{\omega_++B+\Lambda\over 2\omega_+}\ {\omega_-+B-\Lambda\over
2\omega_-}}
\end{equation}
where the roots are non--negative.

Near $p=0$ the energies can be expanded,
\begin{eqnarray}
\omega_\pm &=& (\Lambda \pm B)\sqrt{ 1+{C^2+m^2\over (\Lambda\pm
B)^2}}\nonumber\\
&=&\Lambda\pm B+{1\over 2}\ {C^2+m^2\over \Lambda\pm
B} +\dots\nonumber\\
&=&\Lambda\pm rp^2+{p^2+m^2\over 2\Lambda} +\dots
\end{eqnarray}
for $p,m\ll\Lambda$. In this approximation the eigenspinors (B.3) become
\begin{eqnarray}
u_+ &=&\left( 1+{\gamma_5(ip\!\!\!/ +\phi\cdot T)\over 2\Lambda} -{p^2+m^2\over
8\Lambda^2} +\dots\right)\chi\nonumber\\
&&\nonumber\\
u_- &=&\left({\gamma_5(ip\!\!\!/ +\phi\cdot T)\over \sqrt{p^2+m^2}}
+{\sqrt{p^2+m^2}\over 2\Lambda} +\dots\right)\chi\nonumber\\
&&\nonumber\\
v_+ &=&\left( -{\gamma_5(ip\!\!\!/ +\phi\cdot T)\over \sqrt{p^2+m^2}}
+{\sqrt{p^2+m^2}\over 2\Lambda} +\dots\right)\chi\nonumber\\
&&\nonumber\\
v_- &=&\left( 1-{\gamma_5(ip\!\!\!/ +\phi\cdot T)\over 2\Lambda} -{p^2+m^2\over
8\Lambda^2} +\dots\right)\chi
\end{eqnarray}
and $\beta$ approaches $\pi /2$,
\begin{equation}
\cos\beta ={\sqrt{p^2+m^2}\over\Lambda} +\dots
\end{equation}
To this order the quantities (B.8), (B.9) depend on the barrier height,
$\Lambda$, but not on detailed lattice structure such as the Wilson parameter,
$r$. In the next order such details would begin to appear.

The long wavelength approximation to the propagators $S_\pm$ can be recovered
from (B.7), (B.8)
\begin{eqnarray}
S_\pm (p,t-t') &=& \theta (t-t')\ \sum_\sigma\ u_\pm u^\dagger_\pm\
e^{-i\omega_\pm (t-t')} -\nonumber\\
&&\hspace{3cm} - \theta (t'-t)\ \sum_\sigma\ v_\pm v^\dagger_\pm\
e^{i\omega_\pm (t-t')} \nonumber\\
&=& e^{-i\Lambda\vert t-t'\vert}\left[{1\over 2}\ \varepsilon (t-t')\pm{1\over
2}\ \gamma_5T_c+{\gamma_5(ip\!\!\!/ +\phi\cdot T)\over 2\Lambda} +\dots\right]
\end{eqnarray}
These functions are regular at $p=0$.

The propagator $S_F$ is not regular at $p=\phi =0$. To obtain it one must
consider the 3! possible orderings of the time coordinates, $t,t'$ and $0$,
\begin{eqnarray}
T\left(\psi (p,t)\ \psi (q,t')^\dagger\right)
&=& \theta (t-t')\ \psi (p,t)\ \psi (q,t')^\dagger -\nonumber\\
&&\hspace{2cm} -\theta (t'-t)\ \psi (q,t')^\dagger\ \psi (p,t)\nonumber\\
&=&\theta (t-t')\ \theta (t')\ \psi_+(p,t)\ \psi_+(q,t')^\dagger\nonumber\\
&&+\theta (t)\ \theta (-t')\ \psi_+(p,t)\ \psi_-(q,t')^\dagger\nonumber\\
&&+\theta (-t)\ \theta (t-t')\ \psi_-(p,t)\ \psi_-(q,t')^\dagger\nonumber\\
&&-\theta (t'-t)\ \theta (t)\ \psi_+(q,t')^\dagger\ \psi_+(p,t)\nonumber\\
&&-\theta (t')\ \theta (-t)\ \psi_+(q,t')^\dagger\ \psi_-(p,t)\nonumber\\
&&-\theta (-t')\ \theta (t'-t)\ \psi_-(q,t')^\dagger\ \psi_-(p,t)
\end{eqnarray}
where
\begin{equation}
\psi_\pm (p,t)=e^{-itH_\pm (p)}\ \psi (p)
\end{equation}
The matrix element of (B.11) between free fermion ground states involves the
polarization sums,
\begin{eqnarray}
{\langle +\vert\psi (p)\ \psi (q)^\dagger\vert -\rangle\over\langle +\vert
-\rangle} &=& (2\pi )^4\ \delta_{2\pi} (p-q)\ \sum_\sigma\ {u_+(p,\sigma
)u_-(p,\sigma )^\dagger\over\cos\beta (p,\sigma )}\nonumber\\
{\langle +\vert\psi (q)^\dagger\ \psi (p)\vert -\rangle\over\langle
+\vert -\rangle} &=& (2\pi )^4\ \delta_{2\pi} (p-q)\ \sum_\sigma\ {v_-(p,\sigma
)v_+(p,\sigma )^\dagger\over\cos\beta (p,\sigma )}
\end{eqnarray}
which are obtained by elementary considerations (see Appendix A of Ref.[7])
using the plane wave
expansions,
$$
\psi (p)=\sum_\sigma\left( b_\pm (p,\sigma )\ u_\pm (p,\sigma ) +d^\dagger_\pm
(p,q)\ v_\pm (p,q)\right)
$$
together with canonical anticommutation rules and the ground state definitions
$$
b_\pm (p,q)\vert\pm\rangle =d_\pm (p,\sigma )\vert\pm\rangle =0
$$
On substituting from (B.12) and (B.13) into the ground state matrix element of
(B.11), and using (B.5) to make the time--dependence explicit, one obtains the
expression
\newpage
\begin{eqnarray}
&&S_F(p,t,t') =\nonumber\\
&&=\sum_\sigma\ {1\over\cos\beta}\Bigl[ \theta (t-t')\theta (t')\
e^{-it\omega_+} u_+\left( u^\dagger_+\cos\beta\
e^{it'\omega_+}-v^\dagger_+\sin\beta\ e^{-it'\omega_+}\right)\Bigr.\nonumber\\
&&\quad +\theta (t)\theta (-t')\ e^{-it\omega_+}\ u_+u^\dagger_-\
e^{it'\omega_-}\nonumber\\
&&\quad+\theta (-t)\theta (t-t')\left( e^{-it\omega_-}\cos\beta\
u_-+e^{it\omega_-}\sin\beta\ v_-\right) u^\dagger_-\ e^{it'\omega_-}\nonumber\\
&&\quad-\theta (t'-t)\theta (t)\left(e^{-it\omega_+}\sin\beta\
u_++e^{it\omega_+}\cos\beta\ v_+\right) v^\dagger_+\
e^{-it'\omega_+}\nonumber\\
&&\quad -\theta (t')\theta (-t)\ e^{it\omega_-}\ v_-v^\dagger_+\
e^{-it'\omega_+}\nonumber\\
&&\Bigl.\quad -\theta (-t')\theta (t'-t)\ e^{it\omega_-}\ v_-\left(
-u^\dagger_-\sin\beta\ e^{it'\omega_-}+v^\dagger_-\cos\beta\
e^{-it'\omega_-}\right)\Bigr]
\end{eqnarray}
This function has the expected discontinuity at $t=t'$ but is continuous, as it
should be, at $t=0$ and $t'=0$. Its low momentum behaviour is dominated by the
pole at $\cos\beta =0$. It occurs at $p=\phi =0$ and we can expand around this
point using the formulae (B.8), (B.9). The result is
\begin{eqnarray}
S_F(p,t,t') &=& {1+\gamma_5T_c\over 2}\ {1\over ip\!\!\!/ +\phi\cdot T}\
\gamma_5\ \Lambda\ e^{-i\Lambda (\vert t\vert +\vert t'\vert )}\nonumber\\
&&+{1\over 2}\ \varepsilon (t-t')\ e^{-i\Lambda\vert t-t'\vert}\nonumber\\
&&+i\gamma_5T_c\Bigl\{\Bigl(\theta (t-t')\theta (t')+\theta (-t')\theta
(t'-t)\Bigr) e^{-i\Lambda\vert t\vert}\sin\Lambda t'\Bigr.\nonumber\\
&&\Bigl.\quad +\Bigl(\theta (t'-t)\theta (t) +\theta (-t)\theta (t-t')\Bigr)
e^{-i\Lambda\vert t'\vert}\sin\Lambda t\Bigr\}\nonumber\\
&&+\ {\rm terms\ of\ order}\ \  (p,\phi )
\end{eqnarray}

The poles of $S_F$ are crucial to the continuum limit since they control the
infrared singularities. It is necessary, therefore, to establish conditions
under which the pole at $p=\phi =0$ is unique. According to the general formula
(B.14) the poles of $S_F$ correspond to the zeroes of the function $\cos\beta$
defined by (B.6). In this formula the square roots are non--negative and both
of them must vanish to give a zero of $\cos\beta$. There appear to be two
possibilities,
\begin{equation}
\omega_+=B+\Lambda,\quad\omega_-=-B+\Lambda
\end{equation}
and
\begin{equation}
\omega_+=-B-\Lambda,\quad\omega_-=B-\Lambda
\end{equation}
The alternative (B.17) can be excluded immediately because it implies
$\omega_++\omega_-=-2\Lambda$ which contradicts the positivity of $\omega_+$
and $\omega_-$. The alternative (B.16) is possible only if $\pm B+\Lambda >0$,
i.e. if $B^2<\Lambda^2$. From (B.4) one sees that (B.16) implies $C^2+m^2=0$.
Hence, the zeroes of $\cos\beta$ occur at isolated points defined by
\begin{equation}
C_\mu (p)=0,\quad \phi =0\ \ \ {\rm and}\ \ \
B(p)^2<\Lambda^2
\end{equation}
The origin, $p=0$, is certainly one such point in view of the equations (A.9).
The vector function, $C_\mu (p)$, certainly has other zeroes. This is implied
by the Poincar\'e--Hopf theorem since $C_\mu$ is defined on a torus [20].
In order
that $\cos\beta$ should not vanish at these other points we have only to ensure
that $B(p)^2>\Lambda^2$ at such points. In other words, the zero of $\cos\beta$
at $p=\phi =0$ is unique if $B(p)$ and $C_\mu (p)$ are chosen so as to have no
common zero, apart from the origin, and $\Lambda$ is smaller than $\vert
B(p)\vert$ at all the other zeroes of $C_\mu (p)$.

\newpage

\section*{Appendix C: Time--independent formalism}

In previous work on the overlap prescription we used time--independent
perturbation theory and a different phase convention. The purpose of this
appendix is to clarify the relation between the two formalisms.

With the time--independent approach one constructs the two ground states $\vert
A+\rangle$ and $\vert A-\rangle$ directly by solving the Schroedinger equations
(2.9),
\setcounter{equation}{0}
\renewcommand{\theequation}{C.\arabic{equation}}
\begin{equation}
H_\pm (A)\vert A\pm\rangle =\vert A\pm\rangle E_\pm (A)
\end{equation}
or, rather, the equivalent integral equations
\begin{equation}
\vert A\pm\rangle =\vert\pm\rangle\alpha_\pm +G_\pm (V-E_\pm )\vert A\pm\rangle
\end{equation}
where $\vert\pm\rangle$ denotes the free fermion ground states (2.5). The
operators $G_\pm$ are defined by
\begin{equation}
G_\pm =-{1-\vert\pm\rangle\langle\pm\vert\over H_{0\pm}}
\end{equation}
with the understanding that $G_\pm\vert\pm\rangle =0$. Iteration of (C.2) leads
to the formal solution
\begin{equation}
\vert A\pm\rangle =\alpha_\pm\Bigl( 1-G_\pm (V-E_\pm
)\Bigr)^{-1}\vert\pm\rangle
\end{equation}
where the numerical factors $\alpha_\pm=\langle\pm\vert A\pm\rangle$ are
determined, up to a phase,
by requiring that the states $\vert A\pm\rangle$ be normalized. The energies
$E_\pm$ are determined self--consistently from (C.1),
\begin{eqnarray}
E_\pm &=& {\langle\pm\vert H_\pm (A)\vert A\pm\rangle\over\langle\pm\vert
A\pm\rangle}\nonumber\\
&=& \langle\pm\vert V\Bigl( 1-G_\pm (V-E_\pm )\Bigr)^{-1}\vert\pm\rangle
\end{eqnarray}

The method sketched here is straightforward and practical, at least in the
lowest orders. However, it is less efficient than the time--dependent method
discussed in the text. This is mainly because it requires the calculation of
the subsidiary quantities, $\alpha_\pm$ and $E_\pm$, as well as the quantity of
interest, $\langle A+\vert A-\rangle$. In addition, it obscures the fact that
$\Gamma (A)$ is expressible in terms of connected graphs.

A possible advantage of the time--independent method is the very simple phase
convention it allows. The Brillouin--Wigner convention is expressed in the
requirement that the numerical factors, $\alpha_\pm$, should be real and
positive for all values of the external fields, i.e.
\begin{equation}
\langle\pm\vert A\pm\rangle_{BW}>0
\end{equation}
This makes it easy to compute the angles $\Phi_\pm (\theta ,A)$ induced by
gauge transformations. For example, to first order in $\theta$, the formula
(3.6) reduces to
\begin{eqnarray}
\Phi^{BW}_+ &=& {\rm Re}\left({\langle +\vert F_\theta\vert
A+\rangle\over\langle +\vert A+\rangle}\right)\nonumber\\
&=& {\rm Re}\ \langle +\vert F_\theta\Bigl( 1-G_+(V-E_+)\Bigr)^{-1}\vert
+\rangle
\end{eqnarray}
This was used to compute the chiral anomaly in Refs.[5,11].

To find the relative phase between the B--W ground state and the one used in
the main text one must consider the definitions (2.8). These imply, in
particular,
\begin{eqnarray}
\lim_{\varepsilon\to 0} \ \langle +\vert in+\rangle\ e^{-\hat E_+/i\varepsilon}
&=&\langle +\vert A+\rangle\nonumber\\
&=&\langle +\vert A+\rangle_{BW}\ e^{i\beta_+(A)}
\end{eqnarray}
so that $\beta_+$ can be obtained from the regular part of the connected
amplitude, in the limit $\varepsilon \to 0$,
\begin{equation}
\beta_+(A)=\lim_{\varepsilon\to 0}\ {\rm Im}\left(\langle +\vert
in+\rangle_{con}-{\hat E_+\over i\varepsilon}\right)
\end{equation}
where $\hat E_+$ is itself given by the limit
\begin{equation}
\hat E_+(A)=\lim_{\varepsilon\to 0}\Bigl( i\varepsilon\ \langle +\vert
in+\rangle_{con}\Bigr)
\end{equation}
The contribution of order $N$ to the connected amplitude is expressed as a
1--loop integral constructed with the propagator $S_+$, viz.
\begin{eqnarray}
\langle +\vert in+\rangle_{con} &=& \sum_N\ {(-i)^N\over N!}\int^0_{-\infty}
dt_1\dots dt_N\ \langle +\vert T\Bigl( V(t_1)\dots V(t_N)\Bigr)\vert
+\rangle_{con}\nonumber\\
 &=& -\sum_N\ {(-i)^N\over N}\int^0_{-\infty}
dt_1\dots dt_N\ e^{\varepsilon (t_1+\dots +t_N)}\cdot\nonumber\\
\cdot\int\left({dp_1\over 2\pi}\right)^4\dots
\left({dp_N\over 2\pi}\right)^4&& \hspace{-0.5cm}tr\Bigl(
V(p_1,p_2)S_+(p_2,t_1-t_2)\dots V(p_N,p_1)S_+(p_1,t_N-t_1)\Bigr)\nonumber\\
&&
\end{eqnarray}
It is straightforward to integrate the time coordinates in the terms of this
series using the explicit formula for $S_+$,
\begin{eqnarray*}
S_+(p,t-t') &=&\sum_\sigma\Bigl( \theta (t-t')u_+(p,\sigma
)u_+(p,\sigma)^\dagger\ e^{-i\omega_+(p)(t-t')}-\Bigr.\\
&&\Bigl.\quad -\theta (t'-t)v_+(p,\sigma )v_+(p,\sigma )^\dagger\
e^{i\omega_+(p)(t-t')}\Bigr)
\end{eqnarray*}
One finds, in the limit $\varepsilon\to 0$,
\begin{eqnarray}
\hat E_+(A) &=&\langle +\vert V\vert +\rangle +{1\over 2}\langle +\vert
VG_+V\vert +\rangle +\nonumber\\
&&+{1\over 3}\langle +\vert VG_+VG_+V\vert +\rangle -{1\over 3}\langle +\vert
V\vert +\rangle\langle +\vert VG_+^2V\vert +\rangle +\dots\\
&&\nonumber\\
\beta_+(A) &=& -{1\over 3}\ {\rm Im}\ \langle +\vert VG^2_+VG_+V\vert +\rangle
+\dots
\end{eqnarray}
In the same fashion one can eliminate the time integrations from the terms of
the series (3.10) for $\Phi_+$ to obtain
\begin{eqnarray}
\Phi_+(\theta ,A) &=&\langle +\vert F_\theta\vert +\rangle +{1\over 2}\ \langle
+\vert (F_\theta G_+V+VG_+F_\theta )\vert +\rangle +\nonumber\\
&&+{1\over 3}\langle +\vert (F_\theta G_+VG_+V+VG_+F_\theta
G_+V+VG_+VG_+F_\theta )\vert +\rangle\nonumber\\
&&-{1\over 3}\langle +\vert V\vert +\rangle\langle +\vert (F_\theta
G^2_+V+VG^2_+F_\theta )\vert +\rangle -{1\over 3}\langle +\vert F_\theta \vert
+\rangle\langle +\vert VG^2_+V\vert +\rangle\nonumber\\
&&+{1\over 6}\langle +\vert \Big( [F_\theta ,V]G^2_+V-VG^2_+[F_\theta
,V]\Bigr)\vert +\rangle +\dots
\end{eqnarray}
To the same order the Brillouin--Wigner formula (C.7) gives
\begin{eqnarray}
\Phi_+(\theta ,A)^{BW} &=&\langle +\vert F_\theta\vert +\rangle +{1\over
2}\langle +\vert (F_\theta G_+V+VG_+F_\theta )\vert +\rangle +\nonumber\\
&&+{1\over 2}\langle +\vert (F_\theta G_+VG_+V+VG_+VG_+F_\theta )\vert +\rangle
-\nonumber\\
&& -{1\over 2}\langle +\vert V\vert +\rangle\langle +\vert (F_\theta
G^2_+V+VG^2_+F_\theta )\vert +\rangle +\dots
\end{eqnarray}
These phases are related by
\begin{equation}
\Phi_+(\theta ,A)=\Phi_+(\theta ,A)^{BW}-\delta_\theta\beta_+(A)
\end{equation}
where $\beta_+$ is given by (C.13). This can be verified using the formula
(3.2) or
$$
\delta_\theta V=i[F_\theta ,H_{0+}+V]
$$

According to (C.16) the chiral anomalies are related by
\begin{equation}
\Phi_+-\Phi_-=\Phi^{BW}_+-\Phi^{BW}_--\delta_\theta (\beta_+-\beta_-)
\end{equation}
and we must consider what happens to the functional, $\beta_+-\beta_-$, in the
continuum limit. Like $\Phi_\pm$, the angles $\beta_\pm$ do not involve any
infrared singularities. In the continuum approximation they must be local, i.e.
expressible as integrals over 4--dimensional spacetime of local functions of
the gauge field and its derivatives, with lattice dependent coefficients. More
particularly, the difference, $\beta_+-\beta_-$, must involve a pseudoscalar
density. For example,
$$
\beta_+-\beta_-\simeq{1\over\Lambda^2}\int d^4x\ g^{\rho\sigma}\
\varepsilon^{\kappa\lambda\mu\nu}\ \partial_\kappa\ A_\lambda\ \partial_\rho\
A_\mu\ \partial_\sigma\ A_\nu +\dots
$$
There is no candidate of dimension 4. This means that in the continuum limit,
$k/\Lambda\to 0$, this functional becomes vanishingly small: the continuum
theory chiral anomalies are unaffected by the phase conventions.

\end{document}